\documentstyle[11pt,paspconf]{article} 

\def\edcomment#1{\iffalse\marginpar{\raggedright\sl#1\/}\else\relax\fi}
\reversemarginpar
 
\begin{document}
\title{Martin Schwarzschild's Contributions to Galaxy Dynamics}
\author{David Merritt}
\affil{Department of Physics and Astronomy, Rutgers University, NJ}
 
\section{Introduction}
The astronomical community's debt to Martin Schwarzschild derives from much
more than his published work, as many of us who were his students,
collaborators and friends can testify. Nor did Schwarzschild's contributions
to galaxy dynamics constitute more than a small portion of his scientific
output. Nevertheless it would be hard to think of another single figure whose
work so influenced the development of many of the fields discussed at this
meeting.
 
Those of us who came of scientific age after Schwarzschild's retirement in
1979 tend to identify his contributions to galaxy dynamics with the
remarkable series of papers on elliptical galaxies that began appearing at
about the same time. But Schwarzschild's interest in the structure and
dynamics of stellar systems was lifelong; for instance, as early as 1951, he
published the first of two papers with L. Spitzer concerning the influence of
interstellar clouds on stellar velocities. A number of other papers from this
decade dealt with the relation between the chemical composition and
kinematics of stars in the Milky Way and other galaxies.
 
The following review focusses on three areas of galaxy dynamics where
Schwarzschild's contributions were particularly fundamental: the masses of
stellar systems; the structure of galactic nuclei; and the dynamics of
elliptical galaxies.
 
\section{Masses of Stellar Systems}
The study of the distribution of mass in external galaxies was still in its
infancy when Schwarzschild published his 1954 paper, ``Mass Distribution and
Mass-Luminosity Ratio in Galaxies.'' Here Schwarzschild re-analyzed the
kinematical data in three galaxies -- M31, M33 and NGC 3115 -- for which
earlier workers had found significantly different distributions of light and
mass. In each galaxy, he showed that the data were in fact consistent with a
constant ratio of mass to light, albeit with rather different values in the
three systems. In the case of NGC 3115, for instance, Schwarzschild noted
that a high central velocity dispersion recently measured by Minkowski
implied a large deviation between circular and rotational velocities near the
center of this galaxy, thus allowing $M/L$ to remain approximately constant
in spite of a low central $v_c$.
 
But this paper also contained at least three, quite novel approaches to what
we would now call the ``dark matter problem.'' First, Schwarzschild estimated
the mass of M32 by assuming that its gravitational pull was responsible for
the observed asymmetry in rotation velocity and morphology of its larger
companion M31. He concluded that the mass-to-light ratio of M32 was of order
$200$, in approximate agreement with his value for NGC 3115. Second,
Schwarzschild presented a new and elegant method for evaluating the virial
theorem, the strip-count formula. He showed that the potential energy of a
spherical system could be expressed simply in terms of $S(q)$, the observed
number of objects in a strip of unit width that passes a distance $q$ from
the projected center. \footnote{Strip counts had long been used to infer the
density profiles of star clusters (e.g. Plummer 1911). Schwarzschild was
apparently the first to notice that the potential energy could be computed
directly from $S(q)$ without first converting it into a density profile.} He
applied his technique to the Coma cluster using Zwicky's galaxy counts and
obtained the ``bewilderingly high value'' of 800 for its mass-to-light ratio.
Finally, this paper contained what was probably the first suggestion that
white dwarfs, remnants of an earlier generation of star formation, might
constitute a signficant fraction of the masses of galaxies.
 
In ``Note on the Mass of M92'' (1955), Schwarzschild and S. Bernstein used
the strip-count formula to obtain one of the first accurate measurements of
the mass-to-light ratio of a globular cluster. \footnote{Those familiar with
Schwarzschild's legendary tact will be struck by the introduction to this
paper, which contains a withering (but accurate) critique of a rival formula
for evaluating the virial theorem.}
 
\section{Structure of Galactic Nuclei}
Schwarzschild's pivotal role in the development and deployment of the
balloon-borne telescopes Stratoscope I and II is well known. \footnote{A
wonderfully clear account of the observation of convection cells in the Sun
with Stratoscope I was written by Schwarzschild and his wife, Barbara, for
{\it Scientific American} (1959).} After its two initial flights, Stratoscope
II, a 36-inch telescope, was reconfigured for high-definition photography and
used to obtain images of galactic nuclei unblurred by the atmosphere. In ``An
Upper Limit to the Angular Diameter of the Nucleus of NGC 4151'' (1968,
1973), Schwarzschild, R. Danielson and B. D. Savage reported that the nucleus
of NGC 4151 had still not been resolved and accordingly that only an upper
limit could be placed on its diameter, which they estimated at $0.08''$. They
were thus able to show that the non-thermal continuum, which provides most of
the nuclear light in this Seyfert galaxy, originated in a region much smaller
than that associated with the emission lines.
 
The eighth, and final, flight of Stratoscope II was used to obtain
high-resolution photographs of M31 and M32. The results for M32, while
intriguing, were never published; the observations were made shortly before
sunrise while the telescope was gradually descending and the resultant
temperature differentials caused a substantial degradation in the quality of
the images. But the data seemed to show no evidence for a distinct nucleus at
a resolution of $\sim0.5''$, consistent with what we now know about the
luminosity profile of this galaxy. Observations taken during the same night
of M31 were more successful; in ``The Nucleus of M31'' (1974), E. S. Light,
R. E. Danielson and Schwarzschild presented $0.2''$ resolution photographs
that clearly resolved the nucleus, showing it to have a core radius of only
$0.48''$. More striking was the observed asymmetry of the nucleus, which was
revealed to have a low intensity extension on one side of the bright peak.
Light et al. raised the possibility that the offset was a result of
non-uniform obscuration by dust, and noted that, in the absence of dust,
``the observed asymmetry is an intrinsic property of the nucleus which will
probably require a dynamic explanation.'' The latter picture is now accepted
by most astronomers due to the absence of color variations.
 
\section{Elliptical Galaxy Dynamics}
Starting in 1976, when he was 64 years old, Schwarzschild wrote or
co-authored a remarkable series of 21 papers on the dynamics of elliptical
galaxies. The first of these, a collaboration with M. Ruiz, dates from the
``early days'' of the field when it was still universally assumed that
elliptical galaxies and bulges were rotationally-supported, axisymmetric
systems. ``An Approximate Dynamical Model for Spheroidal Stellar Systems''
(1976) presented a novel approach to the problem of elliptical galaxy
modelling. Ruiz and Schwarzschild wrote $f(E,L_z)=f_0e^{-E/\sigma^2}g(L_z)$,
and assumed in addition that the density generated by $f$ was constant on
spheroids of fixed eccentricity. The two assumptions are mildly inconsistent,
as the authors fully realized, but together they permit an extremely elegant
derivation of the function $g(L_z)$: one first matches the density profile on
the rotation axis, which is independent of $g$, then uses the observed
density in the equatorial plane to determine $g(L_z)$. Ruiz (1976) applied
the model to the central region of M31, treating the nucleus and bulge as
distinct components.
 
The bulge in Ruiz's model of M31 was tipped out of the disk plane in order to
reproduce the observed twist in the isophotes at about $10'$ from the center
of this galaxy. Stark (1977) recognized that a coplanar and triaxial bulge
could reproduce the twist in M31 equally well. At about the same time, a
number of workers began publishing integrated spectra which showed that these
objects were rotating much more slowly than expected for centrifugally
flattened oblate spheroids. Schwarschild contributed to the emerging view of
early-type galaxies as triaxial ellipsoids in two papers with T. B. Williams,
``A Photometric Determination of Twists in Three Early-Type Galaxies,'' I \&
II (1979). These studies revealed significant twists in the inner isophotes
of three elliptical galaxies, which the authors cautiously interpreted as
evidence that ``many elliptical galaxies may have a more complicated basic
structure than that of axially symmetric cofigurations.''
 
Schwarzschild's most famous paper from this period is undoubtedly ``A
Numerical Model for a Triaxial Stellar System in Dynamical Equilibrium''
(1979), in which he constructed the first completely self-consistent model of
a triaxial galaxy. The approach was at the same time beautifully
straightforward and quite novel. Schwarzschild's insight was to treat
individual, time-averaged orbits as building blocks for a galaxy -- thus
replacing the cumbersome self-consistency equations by a matrix equation that
could be solved using standard numerical techniques. In the process, he
discovered the four families of regular orbits in triaxial potentials, the
boxes and the three types of tubes. His demonstration that most orbits in a
non-axisymmetric potential could be regular -- i.e. that they respected three
effective integrals of the motion -- was quite unexpected at the time.
 
Schwarzschild went on, in two subsequent studies, to develop a more complete
understanding of these major orbit families. ``On the Nonexistence of
Three-Dimensional Tube Orbits Around the Intermediate Axis in a Triaxial
Galaxy Model'' (1979), with G. Heiligman, linked the existence of the tube
orbits to the stability of the $1:1$ resonant orbits in the principal planes.
The primary motivation for this work was the apparent absence of
intermediate-axis tube orbits in the self-consistent triaxial model. The
authors showed that the $1:1$ orbit in the $X-Z$ plane \footnote{Here and
below, Schwarzschild's convention is followed in which the $X$ and $Z$ axes
are identified with the long and short axes of the triaxial figure.} (i.e.
the plane perpendicular to the intermediate axis) was generally unstable to
vertical perturbations, a circumstance which they noted was ``quite plausibly
destructive for the existence of $Y$-tube orbits.'' A second study with M.
Vietri, ``Analysis of Box Orbits in a Triaxial Galaxy'' (1983) developed the
picture of box orbits as perturbations of the stable, long-axis orbit. The
key to the analysis was a careful treatment of the second-order terms: these
terms were retained in the development of the transverse motion but omitted
from the axial motion, thus allowing the equations for the different orders
to be solved independently.
 
A remarkable paper from the following year, ``Stellar Orbits in Angle
Variables'' (1984) with S. J. Ratcliff and K. M. Chang, showed how a complete
description of a two-dimensional orbit could be obtained in terms of its
action-angle variables. This problem currently goes under the name of ``torus
construction'' but it is actually quite old, with antecedents in work of
Einstein and Born on semi-classical quantization. Here again, the approach
was beautifully direct. The authors asked simply: How must the Cartesian
coordinates depend on the angles if the angles are to increase linearly with
time? The result was a set of differential equations for $x$ and $y$ as
functions of the angles. These equations are nonlinear, and Ratcliff et al.
developed an iterative technique for solving them which worked well whenever
the initial guess was sufficiently close to the true solution.
 
The slow observed rotation of elliptical galaxies was one of the factors that
prompted Schwarzschild to construct his first triaxial model. Real elliptical
galaxies probably do have rotating figures, and in 1982 Schwarzschild began
investigating the effects of slow figure rotation on the triaxial
self-consistency problem. ``Retrograde Closed Orbits in a Rotating Triaxial
Potential'' (1982), with J. Heisler and D. Merritt, reported the existence of
the ``anomalous'' orbits, $1:1$ resonant orbits that are tipped out of the
$Y-Z$ plane by Coriolis forces. The anomalous orbits give rise to two
families of $X$-tubes that circulate in opposite directions about the long
axis of a rotating triaxial figure. In ``A Model for Elliptical Radio
Galaxies with Dust Lanes'' (1982), T. S. van Albada, C. G. Kotanyi and
Schwarzschild suggested that the dust lanes of Centaurus A and M84 consisted
of matter moving along these anomalous orbits. \footnote{Subsequent
observations of Centaurus A revealed that the sense of rotation of the
stellar body of this galaxy is probably opposite to that of the van Albada
et al. model, implying that the outer dust ring has not yet reached a steady
state. However a triaxial figure is probably still required to support the
inner ring.}
 
Schwarzschild made one attempt at achieving self-consistency in a triaxial
model with rapid figure rotation; this initial attempt failed, as
Schwarzschild reported at one of the Princeton ``Tuesday lunches,'' and the
work was never published. However a subsequent effort, using a more slowly
rotating figure, was successful. In ``Triaxial Equilibrium Models for
Elliptical Galaxies with Slow Figure Rotation'' (1982), Schwarzschild chose a
value for the rotation period that was long enough (of order $10^9$ years
after scaling) that all four of the major orbit families existed out to the
truncation radius of the model. He noted that the two branches of $X$-tubes
must be equally populated if such a model is to be eight-fold symmetric,
which means that a rotating model will lack any streaming around the long
axis. This was another example of how the use of orbits as building blocks
could lead to insights about a galaxy's kinematics that would have been
difficult to obtain from the Jeans or Boltzmann equations.
 
In his 1979 self-consistency study, Schwarzschild had found that box orbits
alone could not reproduce the mass distribution of his triaxial model, since
they tended to place too much mass along the major axis. His solution was to
incorporate $X$-tube orbits which avoid the long axis. Schwarzschild noted
that solutions incorporating the other major orbit family, the $Z$-tubes,
were also likely to exist and that the question of the uniqueness of
solutions ``is thus left unanswered by the present investigation.'' He
returned to the uniqueness question in a 1986 paper, ``Dynamical Models for
Galactic Bars: Truncated Perfect Elliptic Disk.'' Schwarzschild considered a
strongly truncated, planar mass model that supported only one family of
orbits, the boxes, and showed numerically that a self-consistent solution
existed and that it was unique. Beyond the truncation radius in this
two-dimensional model, tube orbits exist in addition to box orbits, and one
might expect to find a certain degree of non-uniqueness in solutions that
draw on both orbit families. This was shown to be the case in a study with P.
T. de Zeeuw and C. Hunter that appeared the following year, ``Nonuniqueness
of Self-Consistent Equilibrium Solutions for the Perfect Elliptic Disk''
(1987). A further step toward demonstrating non-uniqueness in the
three-dimensional problem was taken by Hunter, de Zeeuw, C. Park and
Schwarzschild in ``Prolate Galaxy Models with Thin-Tube Orbits'' (1990). The
authors showed that a variety of self-consistent solutions for axisymmetric
prolate models could be found by varying the relative occupation numbers of
orbits from the two families of thin long-axis tubes.
 
In 1980, R. H. Miller asked Schwarzschild whether he could test the stability
of the nonrotating triaxial model. Schwarzschild agreed, and assigned one of
his students the task of re-integrating the orbits to provide initial
conditions for the $N$-body code. In the process it was discovered that many
of the orbits generated different masses in the grid of cells than they had
in the original integrations. The discrepancy was eventually traced to the
installation of a new computer at the Princeton Computer Center: the
differences in the round-off algorithms of the two machines were sufficient
to trigger the exponential instability of those orbits that were stochastic,
leading to significantly modified trajectories after many orbital periods.
Schwarzschild followed up this hint in the following year in a study with J.
Goodman, ``Semistochastic Orbits in a Triaxial Potential'' (1981). Goodman
and Schwarzschild tested the stability of box orbits by looking for
exponential divergence of nearby trajectories. They noted that a large
fraction of the box orbits were in fact chaotic, but that the chaos produced
only modest changes in the shapes of the orbits over 50 oscillations. They
coined the term ``semi-stochasticity'' to describe this phenomenon. The chaos
was tentatively linked to the linear instability of the short- and
intermediate-axis orbits.
 
Schwarzschild's self-consistent triaxial models from 1979 and 1982 were based
on the Hubble density profile, which has a large, constant-density core. It
became increasingly clear throughout the 1980's that the luminosity profiles
of many galaxies might increase more steeply at small radii; indeed,
Schwarzschild's own Stratoscope observations of M31 and NGC 4151 had revealed
pointlike nuclei in these galaxies. The behavior of box orbits is very
sensitive to the central density of a triaxial model, and in 1989
Schwarzschild began to look in detail at the orbits in triaxial models with
small or nonexistent cores. His two studies with J. Miralda-Escud\'e and J.
F. Lees -- ``On the Orbit Structure of the Logarithmic Potential'' (1989) and
``The Orbital Structure of Galactic Halos'' (1992) -- revealed that the
planar motion in centrally concentrated models is dominated by resonances,
which generate families of orbits not seen in models with large cores.
Schwarzschild, who was fiercely opposed to opaque terminology, gave these
resonant orbits names that evoked their shapes like ``banana,'' ``fish'' and
``pretzel;'' these names have remained in widespread use. He also began to
look in these papers at the behavior of orbits in potentials with central
point masses representing black holes.
 
While Miller \& Smith's (1981) $N$-body study did not find any strong
evidence for instability in Schwarzschild's triaxial model, a number of
examples of dynamical instabilities in other models of hot stellar systems
began to be discussed at about this time. In ``Orbital Contributions to the
Stability of Triaxial Galaxies'' (1989), de Zeeuw and Schwarzschild used an
adiabatic deformation technique to evaluate the stability to small
perturbations of Statler's (1987) triaxial models based on the perfect
ellipsoid. They found that the response of individual box orbits to barlike
perturbations  was often destabilizing, in the sense that the response
density tended to reinforce the original perturbation; a similar mechanism
drives the radial-orbit instability in spherical models. In ``The Ring
Instability in Radially Cold Oblate Models'' (1991), the same authors
investigated axisymmetric instabilities in oblate models constructed from
thin tube orbits. They found that such models were unstable to radial
clumping when sufficiently flat. These stability studies provided yet a
further demonstration of the usefulness of an orbit-based approach to galaxy
dynamics.
 
In one of his last papers, ``Self-Consistent Models for Galactic Halos,''
Schwarzschild revisited the triaxial self-consistency problem, this time
using models based on the singular isothermal mass distribution. Such models
are scale-free, which allowed Schwarzschild to construct orbit libraries by
scaling the orbits computed at a single energy; the increase in efficiency
enabled him to compute orbit libraries for six different choices of the model
axis ratios. Schwarzschild found that most of the box orbits in these models
were significantly stochastic, a rather different situation than he had been
led to expect by his earlier work in two dimensions. He showed that the
omission of the stochastic orbits could sometimes preclude a self-consistent
solution, implying restrictions on the allowed shapes of isothermal halos.
This study demonstrated clearly the importance of chaos in the phase space of
realistic triaxial models and opened to door to a wealth of later studies of
this fascinating topic.
 
\section{Conclusion}
 
It is sometimes said that a scientist's career is over by the age of 35. One
may safely assume that Martin Schwarzschild would have disagreed with this
statement; in any case, all of the work cited here was published after
that particular milestone had been passed. Without the contributions which
Schwarzschild made in the late stages of his career, the field of galaxy
dynamics would be an incomparably less rich and exciting one than it is today.
 
\acknowledgments I am indebted to the following people who provided details 
about Martin Schwarzschild's research or unpublished work, or made helpful 
comments on the manuscript: C. Hunter, R. Miller, F. Schweizer, J. Sellwood, 
T. Statler, P. Teuben, S. Tremaine, T. van Albada, P. Vandervoort, 
T. Williams, and P. T. de Zeeuw.

\end{document}